# Predicting Failure of P2P Lending Platforms through Machine Learning: The Case in China

Jen-Yin Yeh[1], Hsin-Yu Chiu[2,*], and Jhih-Huei Huang[3]

## Abstract

This study employs machine learning models to predict the failure of Peer-to-Peer (P2P) lending platforms, specifically in China. By employing the filter method and wrapper method with forward selection and backward elimination, we establish a rigorous and practical procedure that ensures the robustness and importance of variables in predicting platform failures. The research identifies a set of robust variables that consistently appear in the feature subsets across different selection methods and models, suggesting their reliability and relevance in predicting platform failures. The study highlights that reducing the number of variables in the feature subset leads to an increase in the false acceptance rate while the performance metrics remain stable, with an AUC value of approximately 0.96 and an F1 score of around 0.88. The findings of this research provide significant practical implications for regulatory authorities and investors operating in the Chinese P2P lending industry.

*Corresponding author. E-mail: child300@gmail.com; hychiu@cc.ncu.edu.tw. Address: No. 300, Zhongda Rd., Zhongli District, Taoyuan City 320317, Taiwan (R.O.C.) Tel: +886-3-422-7151 ext. 66533.
[1] Department of Commerce Automation and Management, National Pingtung University. E-mail: jenyiny@mail.nptu.edu.tw.
[2] Department of Information Management, National Central University. E-mail: child300@gmail.com; hychiu@cc.ncu.edu.tw.
[3] Department of Commerce Automation and Management, National Pingtung University. Email: luck99699@gmail.com.

Hsin-Yu Chiu gratefully acknowledges the financial support from the National Science and Technology Council of Taiwan (NSTC111-2410-H-008-071-MY2).

# I. Introduction

The e-finance industry has rapidly grown by integrating Internet technology into traditional financial models, enhancing accessibility to financial services through cost reduction, addressing information gaps, and promoting transparency. Alternative finance markets like P2P lending and crowdfunding challenge traditional banks (Hulme and Wright, 2006; Meyer et al., 2007; Gabor and Brooks, 2017). P2P lending, combining elements of the sharing economy and e-commerce, bypasses traditional banks by enabling direct online lending and borrowing with low minimum loan thresholds and short loan periods (Guo et al., 2016; Ma et al., 2018). Leading international lending platforms employ advanced credit scoring systems and risk management to attract diverse clients. China's P2P finance industry emerged in 2007 with PaiPaiDai's establishment, marking a milestone (Jiang et al., 2018). However, despite rapid growth, the Chinese P2P lending sector encountered issues since 2016 due to the lack of a well-established regulatory system, leading to the closure of numerous platforms.

While many studies explore platform operational efficiency, lending success factors, and overall causes of platform failures (Jiang et al., 2018; Zheng and Wang, 2018; Gao et al., 2018; Chen et al., 2021; Rao, 2021), only a few investigate the connection between platform characteristics and platform survival. Yoon et al.'s (2019) study, pioneering in verifying key factors affecting P2P lending platform default risk, focuses on macro environments, government policies, platform competition, and risk management practices. Gao et al. (2021) utilize transaction data, cash flow, popularity, and platform liquidity as predictors. Both their empirical findings are based on probit models and logistic regressions. Departing from their studies, we examine the predictive power of different types of financial guarantees, various forms of third-party supervision, loan types, and operation models on P2P lending platform survival. Most importantly, this paper identifies relevant variables for directly predicting platform survival and rigorously testing their robustness using various feature selection methods and machine learning models.

We apply six machine learning models: Logistic Regression (LR), Support Vector Machine (SVM), Random Forest (RF), Artificial Neural Network (ANN), Extreme Gradient Boosting (XGBoost), and a Stacking-Based Ensemble Learning model (SBEL) to construct robust prediction models[1]. The SBEL utilizes the other five models as base learners and a logistic regression as the meta learner (final estimator). Beginning with twenty-nine well-documented variables related to P2P lending platform characteristics, we first implement a filter method to select features based on the absolute values of Spearman rank correlation with the target variable. The average performance of the six models is evaluated when the variables are added incrementally. To assess variable importance, we also consider the average performance while adding variables in ascending order of their correlations. We further employ a wrapper method, encompassing forward selection and backward elimination, to select features tailored to each model. To ensure robustness, we also choose exactly five variables using the wrapper method to evaluate the model performance with a compact feature subset, which validates the significance of the selected features.

---

[1] For the inclusion of LR, SVM, RF, and ANN, we follow relevant literature on bankruptcy prediction and credit risks (Kruppa et al., 2013; Tsai et al., 2014; Barboza et al., 2017; Moscatelli et al., 2020; Chen et al., 2023) and employ the most popular supervised learning methods according to the study of Lin et al. (2012). Additionally, we incorporate XGBoost (an ensemble learning method based on boosting) and a stacking-based ensemble learning model (SBEL) to leverage the advantages of ensemble learning.

By analyzing the feature subsets identified across different models and selection methods, we establish a rigorous and practical procedure that ensures the robustness and importance of variables in predicting platform survival. The paper is structured as follows. Section II details data, variable definitions, feature selection methods, machine learning models, and predictive performance measures. Section III presents our research findings. Section IV concludes the study.

## II. Methodology

### Data and Variables

Our analysis focuses on P2P lending platforms from 2014 to the end of September 2018, utilizing data from the websites: Home of Online Lending (wdzj.com) and P2P Eye (p2peye.com). Since late 2014, there have been multibillion-dollar Ponzi schemes disguised as P2P lenders and smaller scams, which have either fled with investor funds or made poor investments. P2P failures surged since mid-2018, leading to a decrease in outstanding loans. For better predictive accuracy, we exclude platforms with less than nine months of regulatory operational experience. The dataset includes 2,438 observations: 1,512 problematic platforms and 926 operational ones[2].

Table 1 outlines variable definitions in detail. The target variable, Operating Status, is 1 for active platforms and 0 for problematic ones. In total, we identify twenty-nine variables: Number of Months of Operation (NoMO), Geographical Location, Registered Capital, Non-state-run Enterprise (NE), Auto Bidding, five indicator variables based on loan types, three indicator variables related to platform operation models, Average Interest Rate (AIR), eight indicator variables based on types of financial guarantees, and seven indicator variables related to third-party supervision. The details in the supporting studies for these variables can be found in Appendix A.

<Table 1 here>

### Feature Selection

We employ feature selection as a pre-processing step to prevent overfitting and reduce computational complexity (Cai et al., 2018). Initially, the filter method is applied to select features independent of machine learning models. We begin with an empty feature set and evaluate the average performance of the six models as variables are added, ordered by descending correlation magnitude. We also assess feature importance by adding variables in ascending order of correlation magnitude. At each step of including a new variable, we tune the hyper-parameters through five-fold cross-validation and randomized searches for each model.

---

[2] More established platforms provide a better basis for understanding default risk factors and building accurate predictive models. The exclusion of platforms operating for less than nine months maintains dataset consistency, allowing for meaningful comparisons with platforms that have a longer operating history. The exclusion is also a practical consideration for investors, helping them avoid very recently established lending platforms. It is also important to note that our dataset is cross-sectional, similar to Yoon et al. (2019) and Gao et al. (2021). Most variables in our study are not time-varying. However, to control for possible time-series variations caused by regulatory policies or other macro variables, we also conduct our analyses with the inclusion of nineteen dummy variables, representing the year-quarter in which a platform defaults. The results show that the significance of variables, optimal feature subsets, and comparison between various models and feature selection methods remain qualitatively consistent. Nevertheless, it is important to note that incorporating these dummies limits the model's applicability to future predictions. We consider these dummies as control variables rather than predictors. Thus, the results obtained by adding the time dummies are available upon request to readers.

Moreover, we utilize the wrapper method to identify the optimal feature subset tailored to each model. Unlike the filter method, the wrapper method selects variables based on the cross-validation score of the AUC, considering interaction effects between non-consecutively ranked variables. We use forward selection and backward elimination (Wang et al., 2016) in the wrapper method. The forward selection starts with an empty subset and gradually adds the variable that contributes the most to the AUC improvement. It stops when no additional variable improves the AUC beyond a predefined threshold. Conversely, backward elimination begins with all variables and removes the one causing the smallest AUC deterioration at each step. The search stops when removing any variable would decrease the AUC by less than a specified negative value. Since the wrapper method needs to assume a set of hyper-parameters to initiate the feature selection process, we tune the hyper-parameters using the training dataset and all twenty-nine variables. After determining the optimal feature subsets, we tune the hyper-parameters again for each model to avoid sub-optimality.

## Machine Learning Models and Performance Measure

Our analysis builds upon previous research and utilizes six models: LR, SVM, RF, ANN, XGBoost, and SBEL. We evaluate the performance using metrics such as Accuracy, Precision, Recall, F1 Score, false acceptance rate (FAR), and area under the receiver operating characteristic curve (AUC). Detailed definitions for the performance measures are provided in Appendix B.

It is important to note that in this paper, a false positive occurs when a platform is predicted to survive but actually fails. Consider a scenario where an investor follows the model's prediction and invests in a platform identified as a survival. In such a case, the investor could potentially lose all invested funds if the platform fails in the future. Conversely, a false negative represents a survival platform incorrectly predicted as problematic. The risk associated with a false negative is that the investor might miss a safe investment opportunity, resulting in a missed chance for profit but no financial loss. Therefore, we emphasize the importance of achieving a lower FAR, as it helps mitigate risk for investors who make investment decisions based on the prediction model.

## III. Research Findings

## Summary Statistics

Table 2 presents the summary statistics for the twenty-nine variables and their Spearman rank correlation with the Operating Status of P2P lending platforms. The average NoMO in our sample is 34.84, with a maximum value of 157. The AIR for the platforms has an average of 12.20%, indicating a relatively high return rate for lenders. The Registered Capital exhibits substantial variability with a standard deviation of 100,746,000. The NoMO and Registered Capital exhibit positive correlations with Operating Status, implying that well-established platforms with longer histories and higher capital are more likely to survive. Conversely, the AIR demonstrates a negative correlation.

<Table 2 here>

Figure 1 presents the variable ranking in descending order of the absolute value of correlation, with the height of the bars indicating the signed numbers. The top five variables are Company License, Bank Deposit Management (BDM), Multiple Loans, NoMO, and No Supervisory Mechanism. Company License shows a positive correlation, indicating third-party approval plays a role in identifying safe or problematic platforms. BDM ranks second, aligning with Boyle et al.'s (2015) argument about countries without deposit insurance being at a higher risk during banking crises. Table 2 reports that 43.07% of the platforms in our sample possess certifications, and 23.17% have entrusted a bank to provide bank deposit services for platform capital. Multiple Loans exhibits a negative correlation, suggesting platforms offering a variety of loans are more likely to be problematic. NoMO strongly correlates with platform sustainability, aligning with Swaminathan's (1996) suggestion of mature firms' survival advantage. The variable No Supervision Mechanism is another highly correlated variable, indicating platforms without proper supervision mechanisms have a higher failure likelihood. Table 2 shows that 79.57% of the platforms lack supervision mechanisms.

<Figure 1 here>

## Feature Selection with Filter Method

Figure 2 illustrates the average predictive performance of the six models as the variables are sequentially added in descending order of correlation magnitude. The graph presents various performance measures: Accuracy, Recall, F1 Score, AUC, and FAR. The vertical axis on the right shows FAR values, while other measures are on the left. The average AUC increases with the number of variables, but the improvement becomes marginal beyond seven variables. Recall and FAR tend to decrease as the number of variables increases but stabilize after the seventh variable. Accuracy and F1 Score changes are subtle as the number of variables increases.

To further demonstrate the feature importance of the variables, Figure 3 displays the average performance of the six machine learning models as the variables are added in ascending order of absolute correlation. Accuracy, AUC, F1 Score, and Recall continue improving as the number of variables increases. Most importantly, the performance is significantly enhanced when adding the last two variables. The FAR values also decrease when the last two variables are included[3].

<Figure 2 here>

<Figure 3 here>

Figures 2 and 3 collectively emphasize that the ranking of correlation in the filter method captures feature importance. As shown in Figure 2, we can arbitrarily select the first seven variables as the feature subset, as more variables do not significantly improve AUC. Although including more variables can further lower FAR, the improvement is minimal.

## Feature Selection with Wrapper Method

Next, we apply the wrapper method to tailor feature subsets specific to the machine learning models. Table 3 outlines the predictive performance of the wrapper method with forward selection. At each stage, the search for the features stops if no variable can improve the cross-validation score of AUC by more than 0.001 (Panel A) or 0.01 (Panel B). The number of variables is smaller in Panel B because the criteria for adding a new variable are stricter. In Panel A, the values of Precision and AUC are higher, while the values of FAR are lower. However, the differences in the AUC are subtle.

---

[3] The value of the FAR is set to 1 when the model predicts no positives. This is because when a model predicts no positives, the value of the FAR would be zero, which can be confused with the situation where the model predicts all survival platforms correctly as positives.

The best-performing model in Panel A, based on AUC, is the XGBoost achieving an AUC of 0.969 and a FAR of 0.065, while the best-performing model in Panel B is the SBEL, with an AUC of 0.961 and a FAR of 0.092. The feature subset for the XGBoost in Panel A consists of six variables, while the SBEL in Panel B utilizes only three variables. Hence, a trade-off exists between variable reduction and predictive performance, as indicated by higher AUC and lower FAR. The differences in other measures are minimal.

<Table 3 here>

Table 3 demonstrates that feature importance is not solely determined by correlation magnitude. While the top three variables (Company License, BDM, and Multiple Loans) with the highest absolute correlation appear in several models, the feature subsets in Panel B also include AIR ranked seventh in Figure 1. In Panel A, Auto Bidding, NE, No Guarantee, and Withdraw Fee are also included in the feature subsets, despite not having the highest rankings based on the absolute correlation. These results highlight the distinction between the filter and the wrapper method.

<Table 4 here>

Table 4 presents the predictive performance of the six models using the wrapper method with backward elimination. The elimination process stops at each stage if removing any variable leads to an adverse change in the cross-validation score of AUC smaller than -0.003 (Panel A) or -0.01 (Panel B).

The best-performing model is the XGBoost in Panel A, with an AUC of 0.968 and a FAR of 0.069. The feature subset consists of seven variables. In Panel B, the same model achieves an AUC of 0.954 with only three variables, but the FAR increases to 0.094, highlighting the trade-off between increasing the number of variables and increasing the FAR.

<Table 5 here>

Finally, we test feature subset robustness by reapplying the wrapper method with forward selection, but this time we limit feature subset size to five variables. The results of the predictive performance are presented in Table 5. The values of the predictive performance measures align with those from Tables 3 and 4. The best-performing model in Table 5 is the SBEL with an AUC of 0.966 and a FAR of 0.076.

Overall, certain variables consistently appear in different models and feature selection methods. Company License and BDM, ranking top two based on Spearman rank correlation, are present in the majority of feature subsets. Multiple Loans and AIR, which negatively correlate with the Operating Status, as well as Auto Bidding and No Guarantee, are also crucial features identified through the wrapper method. These variables demonstrate robustness across models and feature selection methods. Furthermore, the logistic regression models employed by Yoon et al. (2019) and Gao et al. (2021) show effectiveness with hyper-parameter tuning for the inverse of regularization strength and feature selection. However, advanced machine learning models such as XGBoost and SBEL perform better across various feature selection methods.

## IV. Conclusion

In this study, we collect data on Chinese P2P lending platforms and analyze twenty-nine variables using six machine learning models to identify robust features for predicting platform failure. Using filter and wrapper methods with forward selection and backward elimination, we ensure the reliability of our feature subsets. Our findings are as follows. First, incorporating more variables improves predictive performance. However, using the filter method, the improvement in AUC becomes marginal when the feature size exceeds seven, with Accuracy and F1 Score stabilizing around 0.90 and 0.88, respectively. The wrapper method yields AUC and Accuracy values around 0.96 and 0.90.

Second, ranking the absolute correlation is meaningful for selecting important features. The top three variables, i.e., Company License, BDM, and Multiple Loans, appear in the feature subsets across different models. However, variables like Auto Bidding, AIR, Withdrawal Fee, No Guarantee, and NE have lower absolute correlations, but their inclusion decreases the FAR values with the wrapper method.

Lastly, the best-performing model is the XGBoost with six variables using the wrapper method with forward selection, achieving an AUC of 0.969 and the lowest FAR of 0.065. The feature size can be further reduced using different selection methods at the cost of a lower AUC and a higher FAR. Making choices among these models and feature selection methods is equivalent to making decisions regarding more features and higher FAR trade-offs. We leave the decision of selecting the best model to the readers while highlighting the robustness of the identified feature subset across models and selection methods.

The predictive models developed here provide valuable insights for borrowers, investors, and platform operators. Ensuring continuous availability and transparency of platform information is crucial for enhancing operational trajectories and promoting the P2P lending market. While tailored for P2P lending, this paper's variables and predictive models hold relevance for government regulators overseeing digital finance services.

# Appendix A. Supporting Reference of Variables

Previous studies by Yan et al. (2018), Yoon et al. (2019), Chen et al. (2021), and Shao and Bo (2022) emphasize the impact of operation duration and registered capital on platform reputation and survival. Consistent with their findings, we include the variables NoMO and Registered Capital as potential features. Degryse et al. (2018) highlight the significance of a bank's geographical location during financial crises, and Yan (2018) and Chen et al. (2021) utilize the registered city platforms as a relevant variable. Thus, we include the variable Geographical Location to indicate the company headquarters' location, classified into 1st to 5th tier cities based on the official "New Tier City Summit and 2017 China City Business Charisma Ranking".

Following Yan (2018), Chen et al. (2021), and Gao et al. (2021) who investigate the influence of platform ownership, we introduce an indicator variable, Non-state-run Enterprise, denoting whether the platform is government-owned. Jiang et al. (2018) and Shao and Bo (2022) demonstrate the role of automatic bidding mechanisms in mitigating rational herding behavior. Therefore, we incorporate an indicator variable Auto Bidding to denote whether the platform possesses an automatic-bidding mechanism. We employ five indicator variables to categorize the loan types offered by lending platforms. The variety of loan types offered by lending platforms may imply the strategy of business diversification of a platform (Che and Liebenberg, 2017). Furthermore, Emekter et al. (2015), Serrano-Cinca and Gutiérrez-Nieto (2016) and Xia et al. (2017) suggest that default probabilities may vary based on loan purposes.

The platform operation model can be characterized as the financial operations and revenue generation of a platform. We use three indicators to denote whether a handling fee is incurred when borrowers apply for a borrowing plan, when an account is topped up, and when investors withdraw funds from the platform. Additionally, Yan et al. (2018), Yoon et al. (2019), and Shao and Bo (2022) document the significance of the average interest rate of various investments on the platform. Thus, we include the variable AIR as a possible feature. To investigate the risk management strategies of P2P lending platforms, Yan et al. (2018), Yoon et al. (2019), and Chen et al. (2021) include indicator variables to denote whether the platform offers a guarantee contract from a third-party company, whether the platform has a risk reserve account, and whether the cash flow of the platform is managed through a third-party bank. However, the financial guarantee of a platform can take various forms. As illustrated in Table 1, we use eight indicator variables to denote the types of financial guarantees adopted by a platform.

Finally, third-party supervision, which enhances credibility and trust and helps identify and mitigate risks associated with P2P lending, is highly relevant to the survival of P2P lending platforms and to regulatory policies (Wang et al, 2021). However, the existing literature on P2P lending platforms seldom investigates the influence of third-party supervision on platform survival. We apply seven indicator variables to denote the types of third-party supervision that a platform undergoes.

# Appendix B. Performance measure

We define a confusion matrix as Table B.1 where each column of the matrix represents the actual operational state of the sample, and each row of the matrix represents the predicted operational state.

**Table B.1**: Confusion Matrix

| Predicted operational state | Actual operational state | |
|---|---|---|
| | Failure (0) | Survival (1) |
| Failure (0) | True Negatives (TN) | False Negatives (FN) |
| Survival (1) | False Positives (FP) | True Positives (TP) |

TP refers to the number of the survival platforms correctly predicted as safe platforms. TN refers to the number of the platform failures correctly predicted as problematic platforms. FP indicates the number of the platform failures incorrectly predicted as safe platforms. FN indicates the number of the platform survivals incorrectly predicted as problematic platforms.

The Accuracy is then calculated by:

$$Accuracy = \frac{TP + TN}{TP + TN + FP + FN}$$

The precision is defined as:

$$Precision = \frac{TP}{TP + FP}$$

The Recall is calculated as:

$$Recall = \frac{TP}{TP + FN}$$

The F1 Score is defined as:

$$F1\ Score = \frac{2 \times Precision \times Recall}{Precision + recall}$$

The FAR is computed as:

$$FAR = \frac{FP}{FP + TN}$$

The AUC is a metric that quantifies the model's overall performance by measuring the area under the ROC curve (receiver operating characteristic curves). The ROC curve is created by plotting the value of Recall on the y-axis against FAR on the x-axis for different threshold values used to classify samples into positive or negative classes.

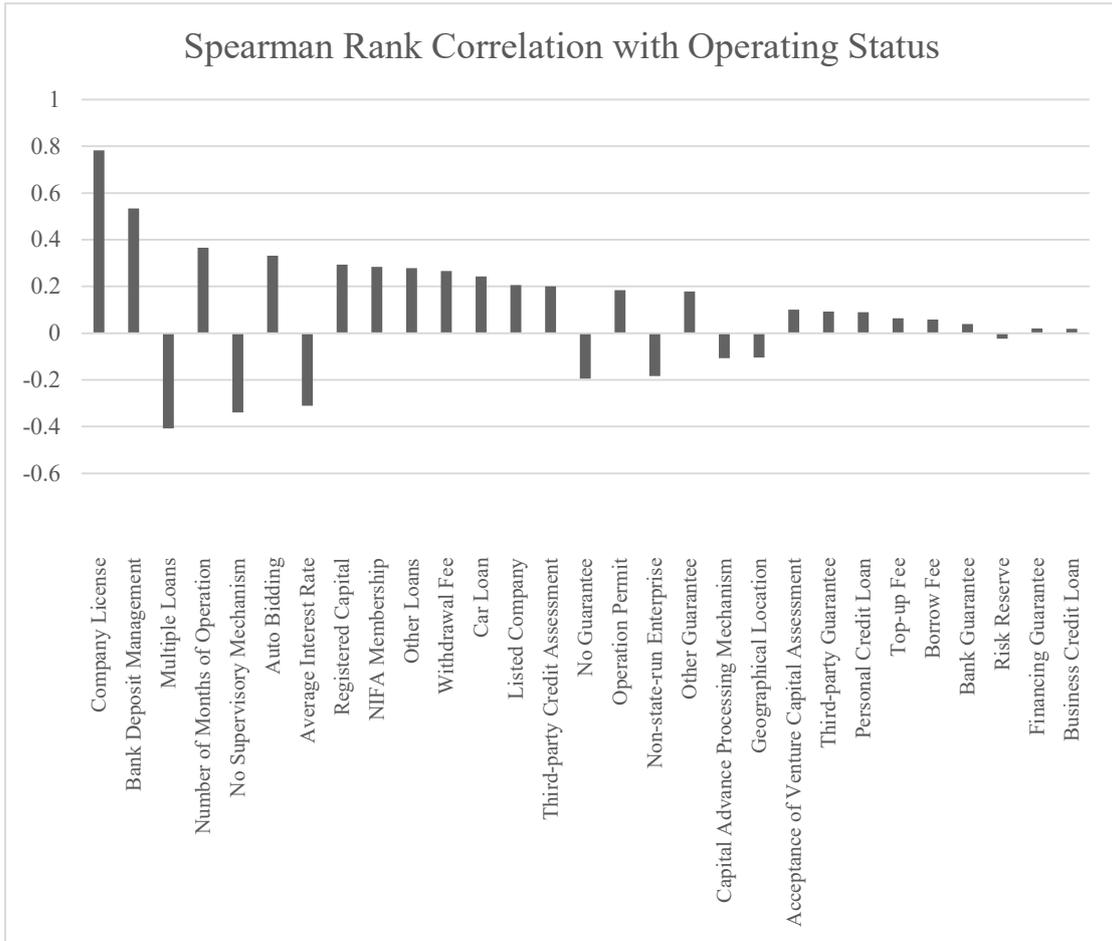

**Figure 1.** Ranking of variables in descending order based on the absolute value of Spearman rank correlation

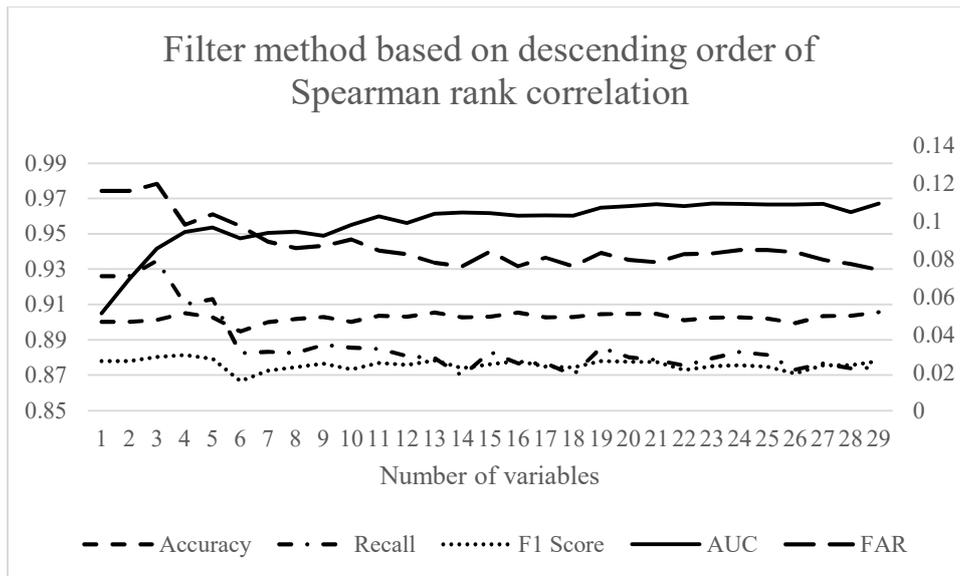

**Figure 2.** Average performance of the six machine learning models with filter method based on descending order of the absolute Spearman rank correlation

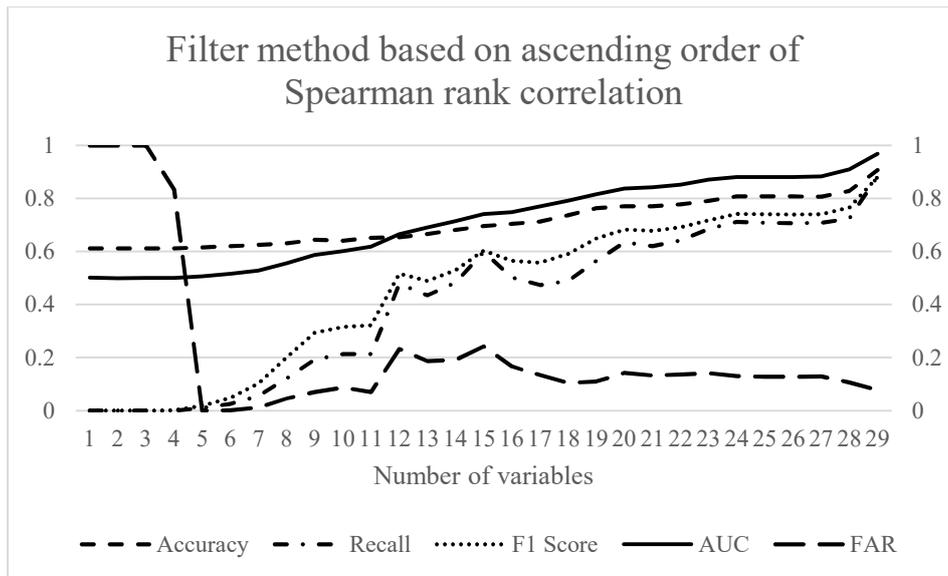

**Figure 3.** Average performance of the six machine learning models with filter method based on ascending order of the absolute Spearman rank correlation

**Table 1.** Variable Definition

| Variable Type | Variable Name | Detail |
|---|---|---|
| Capital and Operation Structure | Target Variable: Operating Status | The platform status is categorized into operating (1) and problematic (0). A platform is labeled as problematic if it experiences making off with the money, termination of business, cash withdrawal failure, or being involved in an investigation. |
| | Number of Months of Operation (NoMO) | The number of months in operation from the platform's launch to the end of September 2018. |
| | Geographical Location | The company headquarters' location is classified into 1st to 5th tier cities based on the official "New Tier City Summit and 2017 China City Business Charisma Ranking". |
| | Registered Capital | The amount of capital registered when the platform was created. |
| | Non-state-run Enterprise (NE) | It equals 0 for government-owned platforms and 1 for non-government-owned platforms. |
| | Auto Bidding | It equals 1 for the platforms that possess an auto-bidding mechanism, and 0 otherwise. |
| Loan Type | Car Loan | It equals 1 for the platforms that offer car loans only and 0 otherwise. |
| | Personal Credit Loan | It equals 1 for the platforms that offer personal credit loans only and 0 otherwise. |
| | Business Credit Loan | It equals 1 for the platforms that offer business credit loans only and 0 otherwise. |
| | Other Loans | It equals 1 if the platform offers a loan type not previously discussed in this study, and 0 otherwise. |
| | Multiple Loans | It equals for the platforms that offer a variety of loan types and 0 otherwise. |
| Platform Operation Model | Borrow Fee | It equals 1 if there is a handling fee incurred when borrowers apply for a borrowing plan on the platform and 0 otherwise. |
| | Top-up Fee | It equals 1 if a handling fee is charged for account top-ups on the platform and 0 otherwise. |
| | Withdrawal Fee | It equals 1 if fees are applied when investors withdraw funds from the platform and 0 otherwise. |
| | Average Interest Rate (AIR) | The average lending rate on the platform for the duration of its operation. |
| Financial Guarantee | Third-party Guarantee | It equals 1 if the platform provides a guarantee contract offered by third-party non-bank institutions for lenders and 0 otherwise. |
| | Bank Guarantee | It equals 1 if the platform provides a guarantee contract offered by third-party banks for lenders, and 0 otherwise. |
| | Risk Reserve | It equals 1 if the platform prepares a risk reserve fund to cover potential losses or defaults and 0 otherwise. |
| | Capital Advance Processing Mechanism (CAPM) | It equals 1 if the platform provides a mechanism to advance capital to borrowers, and 0 otherwise. |
| | Financing Guarantee | It equals 1 if the platform itself offers a guarantee contract for lenders and 0 otherwise. |
| | Bank Deposit Management (BDM) | It equals 1 if the platform has entrusted a bank with bank deposit services for platform capital, and 0 otherwise. |
| | Other Guarantees | It equals 1 if the platform provides guarantee methods not previously discussed in this study, and 0 otherwise. |

**Table 1.** Variable Definition

| Variable Type | Variable Name | Detail |
|---|---|---|
| | No Guarantee | It equals 1 if the platform does not have any guarantee mechanisms, and 0 otherwise. |
| Third-party Supervision | NIFA Membership | It equals 1 if the platform is a member of the National Internet Finance Association of China (NIFA), and 0 otherwise. |
| | Acceptance of Venture Capital Assessment (AVCA) | It equals 1 if the platform has accepted an evaluation from a venture capital firm, and 0 otherwise. |
| | Third-party Credit Assessment (TCA) | It equals 1 if the platform entrusts a third-party credit reporting agency to evaluate the creditworthiness of potential borrowers, and 0 otherwise. |
| | Listed Company | It equals 1 if the platform has been listed on the stock market, and 0 otherwise. |
| | Company License | It equals 1 if the platform obtains a general business license to conduct business activities from the China Banking and Insurance Regulatory Commission (CBIRC), and 0 otherwise. |
| | Operation Permit | It equals 1 if the platform obtains a specialized permit to operate P2P lending transactions from CRIBC, and 0 otherwise. |
| | No Supervisory Mechanism | It equals 1 if the platform does not have any supervisory mechanism, and 0 otherwise. |

**Table 2.** Summary statistics

This table presents the summary statistics of the twenty-nine variables considered as possible features and Operating Status as the target variable. The table provides information on the means, standard deviations, minimum and maximum values, as well as the Spearman rank correlation with Operating Status for Number of Months of Operation, Registered Capital, and Average Interest Rate. For the remaining variables, the table displays the percentages where the variables take a value of 1 and their corresponding Spearman rank correlation with Operating Status.

| Variable | Mean | Stdev | Min | Max | Spearman Rank Correlation with Operating Status |
|---|---|---|---|---|---|
| Number of Months of Operation (NoMO) | 34.84 | 16.37 | 9.00 | 157.00 | 0.3658 |
| Registered Capital | 4,805 | 10,747 | 1 | 285,000 | 0.2934 |
| Average Interest Rate (AIR) | 12.20 | 4.62 | 0 | 48.00 | -0.3106 |

| Variable | Percentage | Spearman Rank Correlation with Operating Status |
|---|---|---|
| Operating Status | 37.98% | -- |
| Non-state-run Enterprise (NE) | 94.67% | -0.1829 |
| Auto Bidding | 16.94% | 0.3315 |
| Car Loan | 6.64% | 0.2425 |
| Personal Credit Loan | 1.97% | 0.0898 |
| Business Credit Loan | 2.34% | 0.0187 |
| Other Loans | 25.88% | 0.2785 |
| Multiple Loans | 62.76% | -0.4075 |
| Borrow Fee | 0.33% | 0.0585 |
| Top-up Fee | 0.94% | 0.0635 |
| Withdrawal Fee | 14.23% | 0.2666 |
| Third-party Guarantee | 9.11% | 0.0931 |
| Bank Guarantee | 0.21% | 0.0392 |
| Risk Reserve | 4.27% | -0.0230 |
| Capital Advance Processing Mechanism (CAPM) | 16.24% | -0.1063 |
| Financing Guarantee | 11.36% | 0.0207 |
| Bank Deposit Management (BDM) | 23.17% | 0.5336 |
| Other Guarantee | 28.10% | 0.1783 |
| No Guarantee | 37.20% | -0.1949 |
| NIFA Membership | 9.60% | 0.2844 |
| Acceptance Venture Capital Assessment (AVCA) | 5.82% | 0.1013 |
| Third-party Credit Assessment (TCA) | 6.07% | 0.2010 |
| Listed Company | 8.53% | 0.2057 |
| Company License | 43.07% | 0.7820 |
| Operation Permit | 7.67% | 0.1841 |
| No Supervisory Mechanism | 79.57% | -0.3393 |

**Table 3** Predictive performance of six machine learning models using wrapper method and forward selection

This table presents the predictive performance of the six models: LR, SVM, RF, ANN, XGBoost, and SBEL. We also provide the feature subsets selected by each model, considering the twenty-nine variables defined in Table 1 are potential features. The predictive performance measures include Accuracy, Precision, Recall, F1 score, FAR, and AUC. We employ the wrapper method to select the feature subsets with forward selection for each model. At each stage, we begin with an empty feature subset and include the best variable that improves the AUC the most, provided the improvement exceeds 0.001 (in Panel A) or 0.01 (in Panel B). If no variable yields an AUC improvement beyond the specified threshold, the feature search stops.

**Panel A: Forward feature selection with a tolerance of 0.001**

| Predictive performance | Models | | | | | |
|---|---|---|---|---|---|---|
| | LR | SVM | RF | ANN | XGBoost | SBEL |
| Accuracy | 0.900 | 0.903 | 0.904 | 0.902 | 0.911 | 0.908 |
| Precision | 0.873 | 0.856 | 0.866 | 0.856 | 0.895 | 0.886 |
| Recall | 0.870 | 0.901 | 0.891 | 0.898 | 0.873 | 0.877 |
| F1 Score | 0.871 | 0.878 | 0.878 | 0.876 | 0.884 | 0.881 |
| FAR | 0.080 | 0.096 | 0.087 | 0.096 | 0.065 | 0.071 |
| AUC | 0.959 | 0.936 | 0.960 | 0.966 | 0.969 | 0.968 |
| Feature Size | 6 | 10 | 8 | 7 | 6 | 8 |
| Feature Subset | Company License, BDM, Multiple Loans, Auto Bidding, AIR, NE | Company License, BDM, Multiple Loans, AIR, Withdrawal Fee, TCA, No Guarantee, Operation Permit, NE, Risk Reserve | Company License, BDM, Multiple Loans, NoMO, Auto Bidding, AIR, Withdrawal Fee, NE | Company License, BDM, Multiple Loans, Auto Bidding, AIR, Withdrawal Fee, AVCA | Company License, BDM, Multiple Loans, AIR, Withdrawal Fee, No Guarantee | Company License, BDM, Multiple Loans, NoMO, Auto Bidding, AIR, Withdrawal Fee, No Guarantee |

**Panel B: Forward feature selection with a tolerance of 0.01**

| Predictive performance | Models | | | | | |
|---|---|---|---|---|---|---|
| | LR | SVM | RF | ANN | XGBoost | SBEL |
| Accuracy | 0.908 | 0.910 | 0.902 | 0.907 | 0.904 | 0.907 |
| Precision | 0.853 | 0.854 | 0.831 | 0.848 | 0.859 | 0.862 |
| Recall | 0.923 | 0.926 | 0.937 | 0.926 | 0.901 | 0.905 |
| F1 Score | 0.887 | 0.889 | 0.881 | 0.886 | 0.880 | 0.883 |
| FAR | 0.100 | 0.100 | 0.121 | 0.105 | 0.094 | 0.092 |
| AUC | 0.956 | 0.922 | 0.951 | 0.957 | 0.954 | 0.961 |
| Feature Size | 3 | 2 | 3 | 3 | 3 | 3 |
| Feature Subset | Company License, Multiple Loans, AIR | Company License, AIR | Company License, BDM, Multiple Loans | Company License, Multiple Loans, AIR | Company License, Multiple Loans, AIR | Company License, Multiple Loans, AIR |

**Table 4** Predictive performance of six machine learning models using wrapper method and backward elimination

This table presents the predictive performance of the six models: LR, SVM, RF, ANN, XGBoost, and SBEL. We also provide the feature subsets selected by each model, considering the twenty-nine variables defined in Table 1 are potential features. The predictive performance measures include Accuracy, Precision, Recall, F1 score, FAR, and AUC. We employ the wrapper method with backward elimination for each model to select the feature subsets. Initially, all twenty-nine variables are included in the feature subset. At each stage, the least important variable is removed if its removal does not significantly impact the AUC by more than -0.003 (in Panel A) or -0.01 (in Panel B). The process is halted if removing any variable results in a negative change in the AUC that is smaller than -0.003 or -0.01.

**Panel A: Backward feature elimination with a tolerance of -0.003**

| Predictive performance | Models | | | | | |
|---|---|---|---|---|---|---|
| | LR | SVM | RF | ANN | XGBoost | SBEL |
| Accuracy | 0.908 | 0.900 | 0.914 | 0.910 | 0.919 | 0.910 |
| Precision | 0.863 | 0.858 | 0.872 | 0.847 | 0.892 | 0.887 |
| Recall | 0.908 | 0.891 | 0.912 | 0.937 | 0.901 | 0.880 |
| F1 Score | 0.885 | 0.874 | 0.892 | 0.890 | 0.897 | 0.883 |
| FAR | 0.092 | 0.094 | 0.085 | 0.107 | 0.069 | 0.071 |
| AUC | 0.959 | 0.941 | 0.967 | 0.959 | 0.968 | 0.965 |
| Feature Size | 5 | 9 | 16 | 5 | 7 | 4 |
| Feature Subset | Company License, BDM, AIR, Other Loans, Car Loan | Company License, BDM, Auto Bidding, AIR, Other Loans, Car loan, No Guarantee, Other Guarantee, Financing Guarantee | Company License, BDM, Multiple Loans, NoMO, AIR, Withdrawal Fee, Car Loan, TCA, Operation Permit, NE, Other Guarantee, CAPM, Geographical Location, Third-party Guarantee, Top-up Fee, Risk Reserve | Company License, BDM, AIR, Other Loans, Car Loan | Company License, BDM, Multiple Loans, AIR, Registered capital, Withdraw Fee, No Guarantee, Location | Company License, BDM, Multiple Loans, AIR |

**Panel B: Backward feature elimination with a tolerance of -0.01**

| Predictive performance | Models | | | | | |
|---|---|---|---|---|---|---|
| | LR | SVM | RF | ANN | XGBoost | SBEL |
| Accuracy | 0.900 | 0.908 | 0.902 | 0.900 | 0.904 | 0.907 |
| Precision | 0.835 | 0.860 | 0.831 | 0.835 | 0.859 | 0.862 |
| Recall | 0.926 | 0.912 | 0.937 | 0.926 | 0.901 | 0.905 |
| F1 Score | 0.878 | 0.885 | 0.881 | 0.878 | 0.880 | 0.883 |
| FAR | 0.116 | 0.094 | 0.121 | 0.116 | 0.094 | 0.092 |
| AUC | 0.945 | 0.918 | 0.951 | 0.950 | 0.954 | 0.961 |
| Feature Size | 3 | 6 | 3 | 3 | 3 | 3 |
| Feature Subset | Company License, BDM, Other Loans | Company License, BDM, Auto Bidding, AIR, Other Loans, Car Loan | Company License, BDM, Multiple Loans | Company License, BDM, AIR | Company License, Multiple Loans, AIR | Company License, Multiple Loans, AIR |

**Table 5** Predictive performance of six machine learning models using wrapper method and forward selection with exactly five variables

This table presents the predictive performance of the six models: LR, SVM, RF, ANN, XGBoost, and SBEL. We also provide the feature subsets selected by each model, considering the twenty-nine variables defined in Table 1 are potential features. The predictive performance measures include Accuracy, Precision, Recall, F1 Score, FAR, and AUC. We employ the wrapper method to select the feature subsets with the forward selection for each model. At each stage, we start with an empty feature subset and include the best variable that improves the AUC the most until the number of variables in the subset exceeds five.

| Predictive performance | Models | | | | | |
|---|---|---|---|---|---|---|
| | LR | SVM | RF | ANN | XGBoost | SBEL |
| Accuracy | 0.900 | 0.900 | 0.906 | 0.893 | 0.895 | 0.910 |
| Precision | 0.870 | 0.835 | 0.846 | 0.860 | 0.903 | 0.884 |
| Recall | 0.873 | 0.926 | 0.926 | 0.866 | 0.817 | 0.884 |
| F1 Score | 0.872 | 0.878 | 0.884 | 0.863 | 0.858 | 0.884 |
| FAR | 0.083 | 0.116 | 0.107 | 0.089 | 0.056 | 0.076 |
| AUC | 0.960 | 0.960 | 0.955 | 0.961 | 0.965 | 0.966 |
| Feature subset | Company License, BDM, Multiple Loans, Auto Bidding, AIR | Company License, BDM, Multiple Loans, AIR, Risk Reserve | Company License, BDM, Multiple Loans, Auto Bidding, Withdrawal Fee | Company License, BDM, Multiple Loans, AIR, Withdrawal Fee | Company License, BDM, Multiple loans, AIR, No Guarantee | Company License, BDM, Multiple loans, AIR, No Guarantee |